\documentclass[aps,twocolumn,prl,amsmath,amssymb]{revtex4-1}
 \usepackage{setspace}
\usepackage{graphicx,epsfig}
\usepackage{color}
\usepackage{wasysym}
\usepackage[normalem]{ulem}
\usepackage{bm}

\begin{document}
\title[]{Haldane and Dimer phases in a frustrated spin chain: an exact groundstate and associated topological phase transition}

\author{Shaon Sahoo$^{1,2,}$}%
%\ead{shaon@iittp.ac.in}%
\author{Dayasindhu Dey$^{1}$}
\author{Sudip Kumar Saha$^{1}$}
\author{Manoranjan Kumar$^{1,}$}
%\ead{manoranjan.kumar@bose.res.in}
\address{$^1$ S. N. Bose National Centre for Basic Sciences, Block JD, Sector III, Salt Lake, Kolkata 700106, India}
\address{$^2$ Department of Physics, Indian Institute of Technology Tirupati, Tirupati 517506, India}

%\date{31/07/2017}

%\parindent = 0pt

\begin{abstract}
A Heisenberg spin-$s$ chain with alternating ferromagnetic ($-J_1^F<0$) and antiferromagnetic ($J_1^A>0$) 
nearest-neighbor (NN) interactions, exhibits the Dimer and spin-$2s$ Haldane phases in the limits $J_1^F/J_1^A \rightarrow 0$ and
$J_1^F/J_1^A \rightarrow \infty$ respectively. These two phases are understood to be topologically equivalent. 
Induction of the frustration through 
the next nearest-neighbor ferromagnetic interaction ($-J_2^F<0$) produces a very rich quantum phase diagram. With frustration, 
the whole phase diagram is divided into a ferromagnetic (FM) and a nonmagnetic (NM) phase. For $s=1/2$, the full NM phase  
is seen to be of Haldane-Dimer type, but for $s>1/2$, a spiral phase comes between the FM and the Haldane-Dimer phases.  
The study of a suitably defined string-order parameter and spin-gap at the phase boundary indicates that the Haldane-Dimer and spiral 
phases have different topological characters. We also find that, along the $J_2^F=\frac 12 J_1^F$ line in the NM phase, an NN dimer 
state is the {\it exact} groundstate, provided $J_1^A>J_C=\kappa J_1^F$ where $\kappa \le s + h$ for applied magnetic field $h$.
Without magnetic field, the position of $J_C$ is on the FM-NM phase boundary when $s=1/2$, but for $s>1/2$, the location of $J_C$ is 
on the phase separation line between the Haldane-Dimer and spiral phases.
\end{abstract}
%\pacs{75.10.Jm, 03.65.Vf}

\maketitle
%\ioptwocol
%%%%%%%%%%%%%%%%%%%%%%%%%%%%%%%%%%%%%%%%%%%%%%%%%%%%%%%%%%%%%%%%%%
%																 %
%		INTRODUCTION											 %
%																 %
%%%%%%%%%%%%%%%%%%%%%%%%%%%%%%%%%%%%%%%%%%%%%%%%%%%%%%%%%%%%%%%%%%

%{\it Introduction.}-- 
\section{Introduction}
\label{sc1}
In condensed matter physics a phase of a system can be identified by defining a suitable order parameter. However, it was later 
found that the phases can not always be characterized by the broken symmetry approach prescribed by Landau; 
here different phases are distinguished according to their topological characters~\cite{chiu16,wen17}.
There are many systems where topology plays a vital role in characterizing their phases~\cite{chiu16,wen17}.
For example, the antiferromagnetic Heisenberg spin chain can have gapped or gapless groundstate depending
on whether site spins are integer or half-integer, as was first conjectured by Haldane~\cite{haldane83}. 
The change in the spectrum of these systems can be explained by the topological terms in the field theoretical description of the  
spin chain \cite{chiu16,haldane83,affleck89}. Later it was shown that the Haldane phase in the odd and even integer spin ($S$) belong 
to different topological classes; while odd-$S$ Haldane phase is topologically nontrivial and protected by symmetries, even-$S$ 
Haldane phase is not protected and can be adiabatically transformed to a trivial site-factorizable phase \cite{pollmann12}. 
It is also worth noting here that, there are topological phase transitions where the symmetry breaking takes place
simultaneouly (e.g. \cite{fujita14}).

Frustrated spin chain models, like `$J_1-J_2$' model, have been extensively studies and they show zoo of quantum 
phases~\cite{ckm,harada,ramasesha95,white_affleck96,mkumar1_10,mkumar2_10,aslam15,aslam16}. In this paper we study a frustrated quantum spin 
chain model, called here AFAF model, which has become a playground for studying and understanding the intricacies of the Haldane phase 
and its associated gap~\cite{hida92a,hida92b,hida92c,hida94}. The AFAF model has Alternating Ferromagnetic ($-J_1^F<0$) 
and AntiFerromagnetic ($J_1^A>0$) nearest-neighbor (NN) exchange interactions. In this model the neighboring spin-$s$ objects 
-connected by ferromagnetic interactions- couple to form an effective spin-2$s$ the Haldane chain in the limit 
$J_1^F/J_1^A\rightarrow \infty$~\cite{hida92a,hida92b,hida92c,hida94,sahoo14}. 
This particular model got more prominence after it was realized experimentally in some materials, e.g.,
$[Cu(TIM)]CuCl_4$ \cite{hagi97}, $(CH_3)_2CHNH_3CuCl_3$ \cite{mana97}, $CuNb_2O_6$ \cite{koda99},
and $(CH3)_2NH_2CuCl_3$~\cite{stone07}.

The main insight gained from studying the $s=1/2$ AFAF model is the topological equivalence of  
the $s=1/2$ the NN dimer phase (in $J_1^F/J_1^A\rightarrow 0$ limit) and  the $s=1$ Haldane gapped 
phase (in $J_1^F/J_1^A \rightarrow \infty$ limit). 
Hida first showed that the Haldane gapped phase is adiabatically connected to the NN dimer state \cite{hida92a} 
by studying the string-order parameter suggested by den Nijs and Rommelse as function of $J_1^F/J_1^A$ \cite{nijs89}.
Kohmoto et al. supported this result by showing that the model has broken $Z_2\times Z_2$ hidden symmetry in both 
the phases \cite{kohmoto92,yamanaka93}.

\begin{figure}[t]
\begin{center} \includegraphics[width=8.6cm,height=4.3cm]{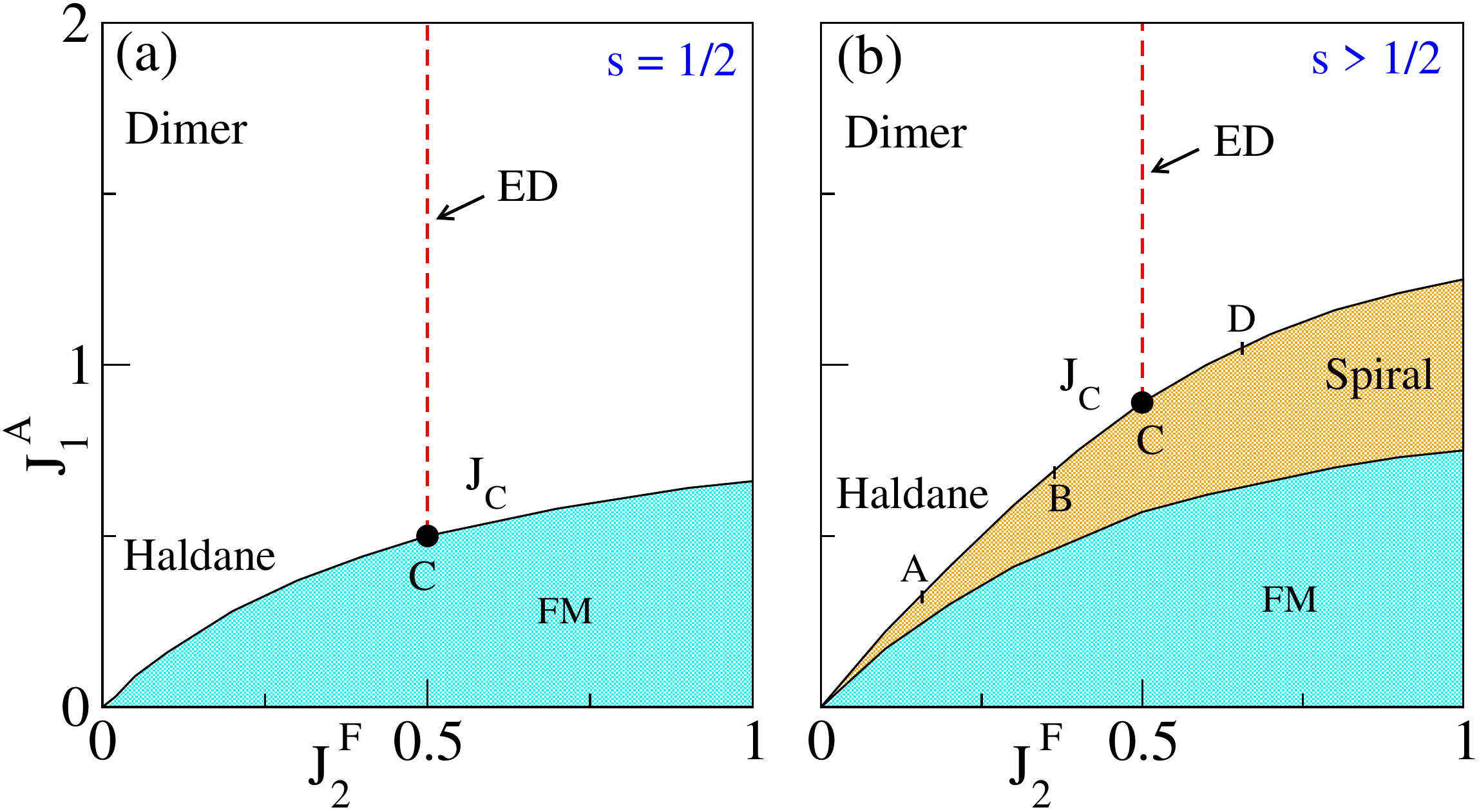}
\caption{Quantum phase diagram of frustrated AFAF spin chain (with $J_1^F=1$). 
For this model, the Haldane and Dimer phases are adiabatically 
connected. (a) $s=1/2$ case: the critical point ($C$) falls in the boundary between the Ferromagnetic (FM) and the Haldane-Dimer 
(HD) phase, (b) $s>1/2$ case: the HD phase is separated from a gapped Spiral phase by a topological phase transition line (ABCD), 
which passes through the point $C$ (location of $C$ depends on $s$). In both the cases, the NN dimer is the exact groundstate along the 
broken line (ED).} 
\label{phsdiag}
\end{center}
\end{figure}

In this paper we study the rich phase diagram of the AFAF spin model with frustration which is induced by the next 
nearest-neighbor (NNN) ferromagnetic interaction ($-J_2^F<0$). With the frustration, the phase diagram shows two main phases 
-ferromagnetic (FM) and nonmagnetic (NM). For $s=1/2$, the whole NM phase is seen to be of gapped Haldane-Dimer type, but for $s>1/2$, 
we also obtain a spiral phase. The study of appropriately defined string-order parameter and spin-gap at phase boundary show us that the 
Haldane-Dimer and the spiral phases are topologically different. We also show here that, an NN dimer state is the {\it exact} groundstate of 
spin model along $J_2^F = \frac 12 J_1^F$ 
line, provided $J_1^A>J_C=\kappa J_1^F$ where $\kappa \le s + h$ for applied magnetic field $h$. While the position of $J_C$ is on the 
FM-NM boundary for $s=1/2$, but for $s>1/2$, it is seen to be on the phase separation line between the topologically distinct Haldane-Dimer 
phase and spiral phases.

This paper is arranged as follows. In the next section (Sec. \ref{sc2}), we explain our spin model. In Sec. \ref{sc3}, we briefly 
discuss equivalence of Dimer and Haldane phases for spin-$s$ AFAF model. Next in Sec. \ref{sc4}, we discuss the phase diagram of the 
frustrated AFAF model -the discussion includes an exact dimer groundstate and the corresponding string-order parameter, and a spiral 
phase for $s>1/2$. In Sec. \ref{sc5}, we discuss a phase transition line (for $s>1/2$) which separates the Haldane-Dimer phase from 
a topologically different spiral phase. We then conclude our paper in Sec. \ref{sc6}.

\section{Frustrated AFAF model} 
\label{sc2}
The following Hamiltonian describes the Frustrated spin-$s$ AFAF chain of size 
$N$:
\begin{eqnarray} 
H ~&=&  ~-~ J_1^F \sum_{k=1}^{N/2} \vec{s}_{2k-1}
\cdot \vec{s}_{2k} ~+~ J_1^A\sum_{k=1}^{N/2} \vec{s}_{2k}
\cdot \vec{s}_{2k+1}\nonumber \\
& & -~ J_2^F \sum_{l=1}^N \vec{s}_{l}\cdot 
\vec{s}_{l+2} - h \sum_{i=1}^N s^z_i, 
\label{fafham} 
\end{eqnarray}
with $J_1^A,~J_1^F,~J_2^F > 0$. Here $\vec{s}_i$'s are site spin
operators with spin value $s$, $J_1^F$ ($J_1^A$) is the NN ferromagnetic (antiferromagnetic)
exchange constant and $J_2^F$ is the ferromagnetic NNN exchange constant. 
The external magnetic field ($h > 0$) is applied along the $z$-direction.
To make sure that the system reduces to an open spin-$2s$ Haldane chain of even number of sites in the
limit $J_1^F/J_1^A \rightarrow \infty$, we only consider here, unless mentioned otherwise, the chain geometry where 
the first and the last bonds are ferromagnetic in nature, and $N\equiv0$ (mod 4).

\section{Equivalence of Dimer and Haldane phases} 
\label{sc3}
In the limit $J_1^F/J_1^A \rightarrow 0$, the groundstate of the Hamiltonian in Eq. \ref{fafham} with $J_2^F=h=0$ is a product of NN 
dimers (here a dimer is the singlet between two spin-$s$ objects).
On the other hand, the model shows a spin-$2s$ Haldane phase in the limit $J_1^F/J_1^A \rightarrow \infty$. 
\begin{figure}[t]
\begin{center} \includegraphics[width=7.6cm,height=3.2cm]{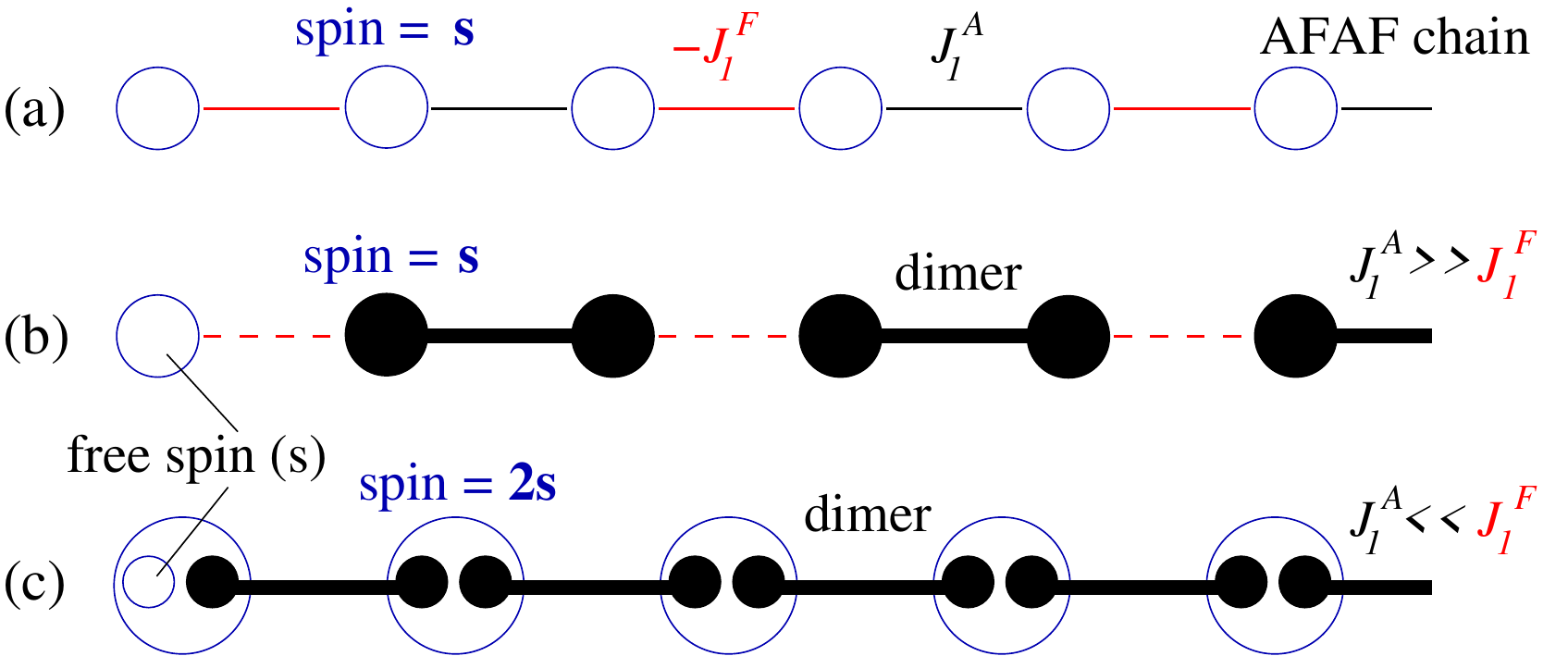}
\caption{(a) Left side of spin-$s$ AFAF chain, (b) The dimer limit ($J_1^F/J_1^A \rightarrow 0$): Two spin-$s$ objects 
are free at the edges, (c) The Haldane limit ($J_1^F/J_1^A \rightarrow \infty$): after formation of VBS state, there will be two 
free spin-$s$ objects left at the edges.} 
\label{dimhal}
\end{center}
\end{figure}
For $s=1/2$, the Dimer and Haldane phases are known to be adiabatically connected and topologically equivalent 
\cite{hida92a,kohmoto92,yamanaka93}. But this equivalence result is expected to be valid even for $s>1/2$. This can be understood 
in the following way. Since different topological states can be distinguished by their groundstate degeneracies \cite{wen90,wen17}, 
we will now calculate this quantity in two opposite limits. We first note 
that, in the dimer limit ($J_1^F/J_1^A \rightarrow 0$), there exist two free spin-$s$ objects at the
two edges of the chain (see FIG. \ref{dimhal}b); this gives rise to the $(2s+1)$-fold groundstate degeneracy. 
In the Haldane limit ($J_1^F/J_1^A \rightarrow \infty$), we  get a spin-$2s$ open chain of length $N/2$. 
The groundstate of this Haldane chain is actually topologically equivalent to the valance bond solid (VBS) state.  
The VBS state for spin-$2s$ system can be obtained in the following way: each site spin can be considered as symmetric combination 
of two spin-$s$ objects \cite{aklt87}. Now singlets can be formed between two spin-$s$ objects from two neighboring sites 
(see FIG. \ref{dimhal}c). This leaves two free spin-$s$ objects at the two edges of the chain \cite{pollmann12,jiang10}.
This shows that, in the Haldane limit too, the groundstate degeneracy will be $(2s+1)^2$. 
The same value of the groundstate degeneracy in two opposite limits 
indicates that, the Dimer and Haldane phases are topologically equivalent for all $s$. 

It may be mentioned here that, for the odd-spin Haldane phase, 4-fold degeneracy is genuine (as long as the time-reversal
symmetry is maintained) and the rest of the degeneracy can be lifted by perturbations. On the other hand, for the even-spin Haldane phase, 
full degeneracy is accidental and can be lifted by perturbations \cite{pollmann12}. The preceding discussion is valid in both the cases in the 
sense that it only shows the equivalence of the Haldane phase and the Dimer phase, irrespective of whether those phases are topological 
trivial or not.

%% EXTENDED PHASE DIAGRAM 

\section{Extended phase diagram with frustration}
\label{sc4}
To better understand the behavior of the Haldane and Dimer phases under the frustration, and to investigate other possible 
quantum phases, we now study the AFAF model with frustration induced by the NNN ferromagnetic interactions ($-J_2^F<0$). In this section, we
first show that an NN dimer state is the {\it exact} groundstate of the Hamiltonian in Eq.~\ref{fafham} along a special line in the 
parameter space. Then we exactly calculate a suitably defined string-order parameter and show that how our spin system behaves differently 
for integer or half-integer spin $s$. In the next two sub-sections, we respectively discuss a spiral phase ($s>1/2$) and a ferromagnetic phase 
that appear in the phase diagram for $J_2^F>0$.

\subsection{Exact ground state}
We begin by showing 
that, when $J_2^F = 1/2$ (we set $J_1^F=1$ as normalization), an NN dimer state is an {\it exact} eigenstate of the Hamiltonian 
in Eq.~\ref{fafham}. This dimer state is then proved to be the exact groundstate when $J_1^A$ is larger than a critical value. 
Along the special line $J_2^F = 1/2$ in the parameter space, we rewrite the Hamiltonian in the following form (assume 
periodic boundary condition):
\begin{eqnarray} 
H~=~J_1^A \sum_{k=1}^{N/2}~ \vec{s}_{2k}\cdot 
\vec{s}_{2k+1}-\frac 12 \sum_{k=1}^{N/2}~\vec{s}_{2k}
\cdot (\vec{s}_{2k-2} + \vec{s}_{2k-1}) &\nonumber \\ 
~- \frac 12 \sum_{k=1}^{N/2}~\vec{s}_{2k-1}\cdot 
(\vec{s}_{2k} + \vec{s}_{2k+1}) 
~-h\sum_{k=1}^{N/2} (s^z_{2k}+
s^z_{2k+1}).~~~~ &
\label{splham} 
\end{eqnarray}
Let $[i,j]$ be the singlet state between spins at sites $i$ and $j$. We then
have the following relations: $\vec{s}_i \cdot \vec{s}_j 
[i,j] = -s(s+1)[i,j]$, $\vec{s}_k \cdot(\vec{s}_i +
\vec{s}_j)[i,j] = 0$ for all $k~ \ne i, j$ and
($s^z_i + s^z_j)[i,j] = 0$.
Using these relations, it is easy to verify that the state $\psi = 
[2,3][4,5][6,7] \cdots [N,1]$ is an eigenstate of the Hamiltonian
$H$ with the eigen energy $E_d=-\frac N2 s(s+1)J_1^A$, i.e.,
\begin{eqnarray} 
H~\psi &=& -\frac N2 s(s+1)J_1^A~\psi.
\label{egnst} 
\end{eqnarray}
Using the Rayleigh-Ritz variational principle, it is possible to prove that this $\psi$ is a groundstate of the system
when $J_1^A$ is greater than a critical value $\kappa$ ($=J_C$ with $J_1^F=1$). In the following we show that 
$\kappa \le s + h$.

Suppose that the total Hamiltonian of
a system is written as the sum of $M$ terms, i.e., $H = \sum_{i=1}^M H_i$. Using the
Rayleigh-Ritz variational principle, it can be shown that if a state is
simultaneously a groundstate of each of $H_i$'s, then it will also
be a groundstate of the total Hamiltonian. To use this theorem for our purpose, 
we decompose $H$ in Eq. \ref{splham} as $H = \sum_{k=1}^{N/2} (H_{2k-1} ~+~ H_{2k})$, where
\begin{eqnarray} 
H_{2k} = \frac 12 J_1^A~ \vec{s}_{2k-2}\cdot \vec{s}_{2k-1} - \frac 12 \vec{s}_{2k} \cdot 
(\vec{s}_{2k-2} + \vec{s}_{2k-1})\nonumber\\
-\frac h2 (s^z_{2k-2}+s^z_{2k-1})~~ {\rm and}\\
H_{2k-1} = \frac 12 J_1^A~\vec{s}_{2k}\cdot \vec{s}_{2k+1} - 
\frac12 \vec{s}_{2k-1}\cdot (\vec{s}_{2k} + \vec{s}_{2k+1}) \nonumber\\
-\frac h2 (s^z_{2k}+s^z_{2k+1}).
\label{hblocks} 
\end{eqnarray}
Here each part, represented by either $H_{2k}$ or $H_{2k-1}$, corresponds to a block of three spins. All these block Hamiltonians 
are essentially equivalent and have the same eigenvalues. Let us denote the Hamiltonian for this three-spin
block as $H_{u}$; considering the first triangle, we have
$H_{u}=\frac 12 J_1^A~\vec{s}_{2}\cdot \vec{s}_{3} -
\frac12 \vec{s}_{1}\cdot (\vec{s}_{2} + \vec{s}_{3})-\frac h2 (s^z_{2}+s^z_{3})$.
In an earlier work \cite{sahoo14}, it was shown by explicitly forming and diagonalizing
$H_{u}$ (with $h=0$) that, the NN dimer state $\psi$ is the groundstate of $H$ for $J_1^A\ge0.5$ when
$s=1/2$ and for $J_1^A\ge1.0$ when $s=1$. Numerically $\kappa$ value is found to be 0.5 and 0.9 respectively
for $s=1/2$ and $s=1$. We now extend this result for general spin $s$, in addition, this time we
also consider the presence of (weak) magnetic field $h$.

First we rewrite $H_{u}$ in the following way:
\begin{eqnarray} 
H_{u} &=& -\frac 14 (\vec{s}_1+\vec{s}_2+\vec{s}_3)^2+\frac 14 (J_1^A+1)(\vec{s}_2+\vec{s}_3)^2 
\nonumber \\
&& - \frac h2 (s^z_2+s^z_3)+\frac 12 (\frac 12 -J_1^A)s(s+1).
\label{hblck} 
\end{eqnarray}
The first two terms in Eq.~\ref{hblck} compete with each other and determine
the nature of the groundstate in spin block. For example, if $J_1^A \to 0$, all three
spins will try to form maximum possible total spin $3s$ and
minimize the energy of $H_{u}$. On the other hand,
if $J_1^A \gg 0$, there will be a singlet formation between 2nd and 3rd spins to minimize the value of
the 2nd term in the expression of $H_{u}$. Therefore, in the large $J_1^A$ limit and in the presence
of a weak magnetic field,
the lowest energy of a three-spin block would be $E_s=-\frac 14 s(s+1) + \frac 12 (\frac 12 - J_1^A)s(s+1)
=-\frac {J_1^A}{2}s(s+1)$. For $N$ such singlet state of three-spin
blocks, the total energy for the whole system would be
$-\frac{N}{2} J_1^A s(s+1)$ which have the value same as the eigen energy of the NN dimer state $\psi$
 in Eq. \ref{egnst}. Therefore, following the theorem stated earlier,
$\psi$ will be the groundstate of the total system in the large $J_1^A$ limit.

Now we try to estimate the critical value of $J_1^A$, denoted here by $J_C$, above which $\psi$ is guaranteed to be the 
groundstate. In the phase diagram FIG. \ref{phsdiag}, the corresponding point is marked by $C$, and $ED$ marks a line 
along which $\psi$ is an exact groundstate. For the estimation of $J_C$, we first note the
following tendency of a spin block. Let us first start with a large value of $J_1^A$ for which a singlet is formed to
minimize the energy of a three-spin block. Now as we reduce the value of $J_1^A$, the three-spin block will try to
minimize its energy by increasing the absolute value of the 1st term and lowering the value of the 2nd term (see
Eq. \ref{hblck}). At the transition point, the singlet between the 2nd and the 3rd site would break and a triplet would
be formed. In addition, the three spins together will form a spin $(s+1)$ to maximize the absolute value of the
1st term. Correspondingly, the energy of the spin block would be
$E_t=- \frac 14 (s+1)(s+2)+\frac 14 (J_1^A+1) 1(1+1) -\frac h2 (1) +\frac 12 (\frac 12 -J_1^A)s(s+1)
=-\frac {J_1^A}{2}s(s+1) + \frac 12 (J_1^A -s - h)$. The NN dimer state $\psi$ will be the groundstate
of the Hamiltonian in Eq. \ref{splham} as long as $E_s\le E_t$, i.e., when $J_1^A \ge s + h$. This gives us the
upper bound of the critical value $\kappa$, above which $\psi$ is guaranteed to be the groundstate; so we have
$\kappa \le s + h$.

\begin{figure}[t]
%\begin{center} \includegraphics[width=8.3cm,height=4.4cm]{spgap_strord}
\begin{center} \includegraphics[width=\columnwidth]{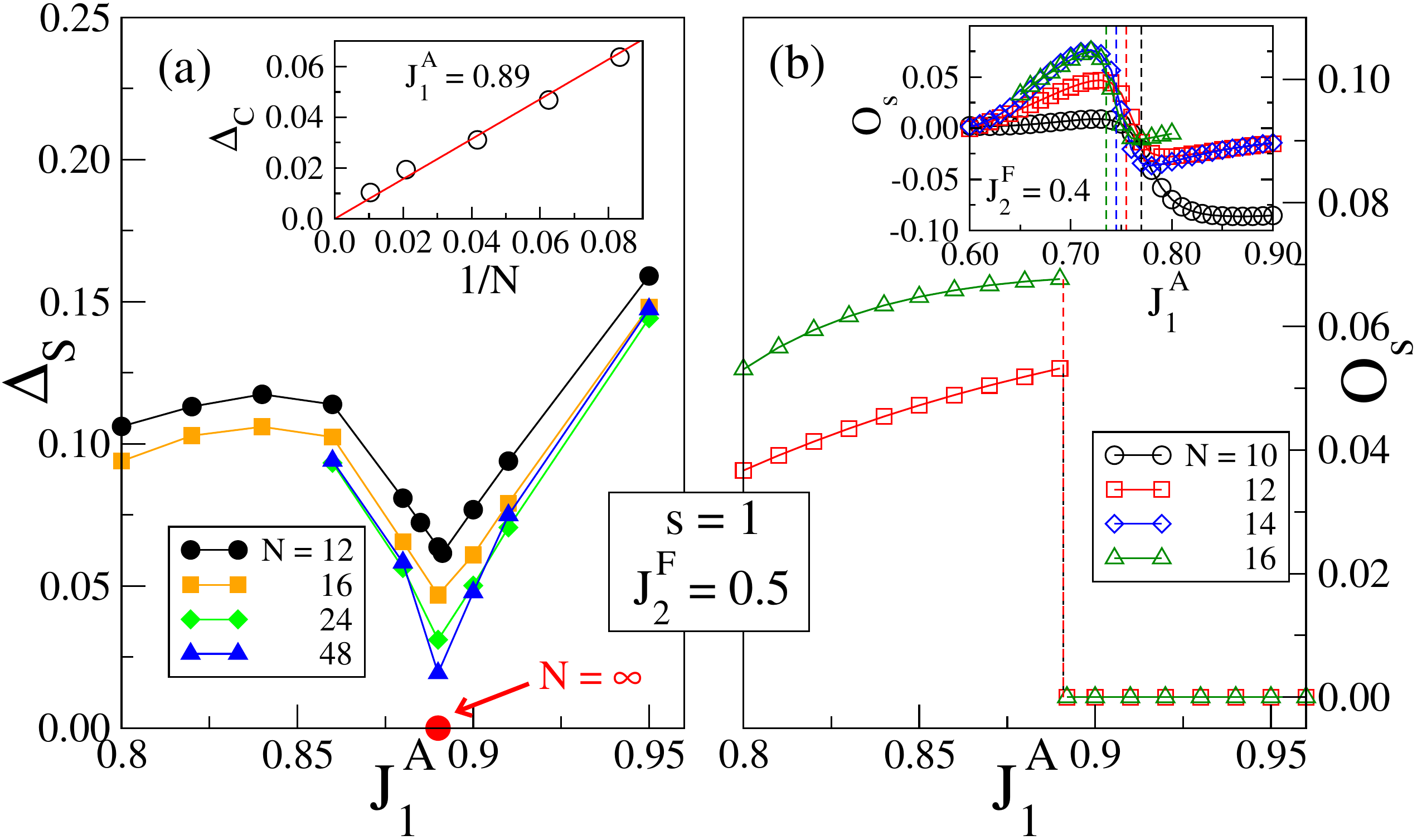}
\caption{Numerical results for $s=1$ frustrated AFAF closed chain. (a) Spin-gap ($\Delta_S$) at the critical point $C$ 
($J_2^F=0.5$, $J_1^A\approx0.89$) decreases with system size (N), and expected to go to zero as $N\rightarrow \infty$ (inset). 
In the inset, we have an additional point corresponding to N = 96. (b) String-order parameter ($O_S$) shows sudden change across the point 
$C$. $O_S$ also changes across any point on the ABCD line (see FIG. \ref{phsdiag}). The change becomes sharper with increasing system size 
(in the inset, results shown for one representative point $J_2^F=0.4$). Broken lines are for eye-guides, indicating where change in $O_S$ 
is large.} 
\label{spn1plts}
\end{center}
\end{figure}

\begin{figure}
%\begin{center} \includegraphics[width=8.3cm,height=4.4cm]{spin_3by2_spin_half}
\begin{center} \includegraphics[width=\columnwidth]{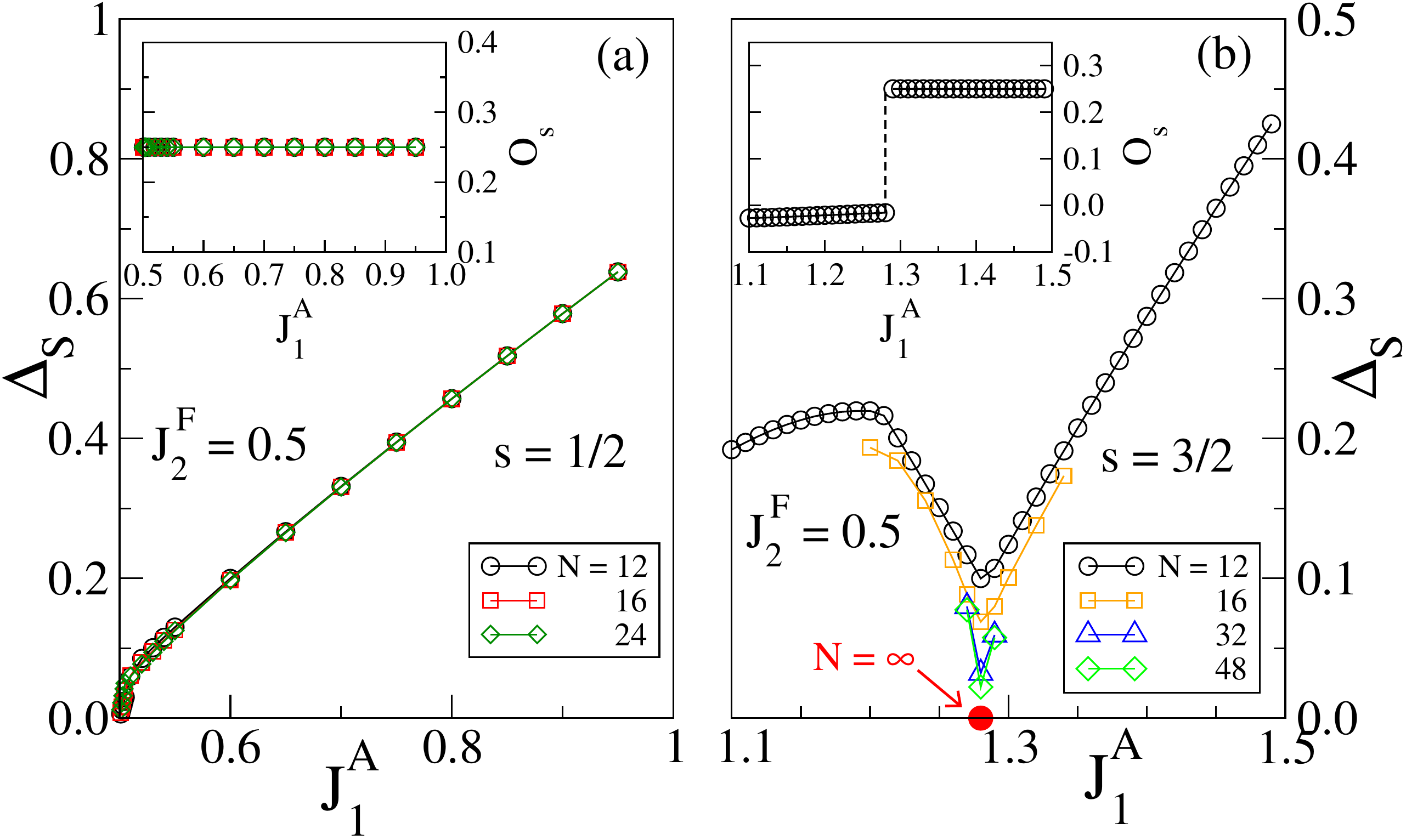}
\caption{(a) $s=1/2$ case: Spin-gap ($\Delta_S$) is seen to go to zero at point $C$ ($J_2^F=0.5$,$J_1^A=0.5$). 
String-order parameter ($O_S$) is always 1/4 in for the
dimer groundstate (inset). (b) $s=3/2$ case: $\Delta_S$ goes through a minimum at point $C$ ($J_2^F=0.5$,$J_1^A\approx1.28$); 
this minimum is expected to go to zero with system 
size (N). $O_S$ goes though a sudden change across the point $C$ and it takes the value 1/4 in the dimer groundstate (inset).}
\label{spn_intby2}
\end{center}
\end{figure}

To understand the character the point $C$, we note for $s=1/2$ that, this point lies in the ferromagnetic-nonmagnetic 
transition line. For $s>1/2$, this point lies inside the nonmagnetic region and the spin gap ($\Delta_S$ = singlet-triplet gap) at 
this point is found to be zero within our numerical accuracy (more details on the ferromagnetic and nonmagnetic phases are given later). 
The relevant results on the spin gap for $s=1$ and $s=3/2$ can be found in FIGs. \ref{spn1plts}a and \ref{spn_intby2}b respectively. 
As the gap is expected to be zero at point $C$, we call this point a critical point. 

It is interesting to note here that, the NN dimer state is the exact groundstate of the spin model in two different limits of 
the extended phase diagram -- along the line $J_2^F=\frac 12$ with $J_1^A>J_C$, and when $J_2^F=0$ with 
$J_1^F/J_1^A \rightarrow 0$.
The study of the spin-gap, string-order parameter and the entanglement spectrum for $s=1/2$ and 1 indicates 
that the dimer phases in those two limits are adiabatically connected and one does not encounter a quantum phase transition while 
going from one exact dimer groundstate to the other \cite{sahoo14}. We expect this to be true for all values of $s$. 
The adiabatic connectivity explains the continuity of the upper part of the phase diagrams in FIG. \ref{phsdiag}.

\subsection{The string-order for dimer state}
The topological equivalence between the Dimer and the Haldane phases, as discussed in Sec. \ref{sc3}, allows us to gain more insight into 
the nature of the Haldane phase by studying the topological character of the exact NN dimer groundstate. Here we first analytically 
calculate the string-order parameter ($O_S$) for the exact dimer state.  
For our AFAF spin model, the string-order parameter is defined as: $O_S=\lim_{|k-l|\rightarrow \infty}O_{kl}$, where 
\begin{eqnarray} 
O_{kl}=- \langle (s^z_{2k-1}+s^z_{2k}) \exp\{i\pi\theta_{kl}\}(s^z_{2l-1}+s^z_{2l}) \rangle,
\label{sopmr}
\end{eqnarray}
with $\theta_{kl}=s^z_{2k+1}+s^z_{2k+2}\cdots+s^z_{2l-2}$. We define $O_{kl}$ in such a way that, both $(\vec{s}_{2k-1}+\vec{s}_{2k})$ and 
$(\vec{s}_{2l-1}+\vec{s}_{2l})$ form a spin-$2s$ object in the Haldane limit $J_1^F/J_1^A \rightarrow \infty$. Soon we will 
find that this defination of $O_S$ produces result which is consistent with the earlier result that the odd and even spin Haldane phases are 
topologically nontrivial and trivial respectively.

The singlet (dimer) state between two spin-$s$ objects is 
given by $|d \rangle =\sum_{m=-s}^s \frac{(-1)^{s+m}}{\sqrt{2s+1}} |-m,m \rangle$; here basis $|a,b \rangle$ denotes a state where 
$z$-component of the first (second) spin is $a$ ($b$). The dimer groundstate of the spin-$s$ system can now be written as:
$|\psi\rangle=\Pi_{k=1}^{\frac N2}|d_{2k,2k+1}\rangle$, where $|d_{2k,2k+1}\rangle$ denotes the singlet state between sites $2k$ and $2k+1$. 
While evaluating $O_{kl}$, we will assume that $|k-l|>1$. Since $(s^z_{2k}+s^z_{2k+1})|d_{2k,2k+1}\rangle=0$, it is 
evident that, $\exp\{i\pi\theta_{kl}\}=\exp\{i\pi(s^z_{2k+1}+s^z_{2l-2})\}$ for the dimer groundstate $|\psi\rangle$. 
We also note that, due to the special form of the groundstate, $\langle s^z_{2k-1}\exp\{i\pi\theta_{kl}\}s^z_{2l} \rangle$ 
= $\langle s^z_{2k-1} \exp\{i\pi\theta_{kl}\}s^z_{2l-1} \rangle$ = $\langle s^z_{2k}\exp\{i\pi\theta_{kl}\}s^z_{2l}\rangle$ = 0.  
With this information, $O_{kl}$ can now be written in the following form: $O_{kl}=-L \times R$, where 
$L=\langle d_{2k,2k+1}|s_{2k}\exp\{i\pi s^z_{2k+1}\}|d_{2k,2k+1}\rangle$ and 
$R=\langle d_{2l-2,2l-1}|\exp\{i\pi s^z_{2l-2}\}s^z_{2l-1}|d_{2l-2,2l-1}\rangle$. We note that the values of $L$ and $R$ are the same, and 
equal to $\langle d_{2,3}|s_{2}\exp\{i\pi s^z_{3}\}|d_{2,3}\rangle$. Therefore we can write,
\begin{eqnarray} 
O_{kl}&=&- \left[\langle d_{2,3}|s_{2}\exp\{i\pi s^z_{3}\}|d_{2,3}\rangle\right]^2 \nonumber \\
      &=&-\frac {1}{(2s+1)^2}\left[\sum_{m=-s}^{s}m~e^{i\pi m}\right]^2 \nonumber \\
      &=&\frac {4}{(2s+1)^2} \left[\sum_{m=a}^{s}m~sin (\pi m) \right]^2,
\label{sofnl}
\end{eqnarray}
where $a$ = 1 or 1/2 depending on whether $s$ is integer or half-integer respectively. After doing some algebra, we find from 
Eq. \ref{sofnl} that, $O_{kl}$ = 0 or 1/4 when $s$ is integer or half-integer. This simply shows that, the value of $O_S$ for the 
NN dimer state is 0 or 1/4 depending upon whether $s$ is integer or half-integer.

The odd spin Haldane phase is a Symmetry Protected Topological or SPT phase \cite{gu09}, on the other hand, the even spin
Haldane phase is a trivial phase \cite{pollmann12}. Since the spin-$2s$ Haldane phase is topologically equivalent to the Dimer phase of 
spin-$s$ AFAF model, we infer from the above result that the string-order parameter, as defined in Eq. \ref{sopmr}, accurately identifies 
for our spin model whether a phase has topological character or not. We therefore use this suitably defined string-order parameter (as well 
as the spin-gap) to study different topological phases and associated topological phase transitions for our spin model (see Sec. \ref{sc5}).

\subsection{Ferromagnetic phase}
The extended phase diagram of the frustrated AFAF model, as shown in FIG. \ref{phsdiag}, consists of a ferromagnetic region (FM) and 
nonmagnetic region (NM). The FM region in the diagram enlarges as the ferromagnetic NNN interaction ($-J_2^F$) increases in strength. 
A classical analysis \cite{hida13,sahoo14}, as well as a spin-wave analysis \cite{sahoo14}, finds that the phase boundary between them is 
determined by $J_2^F=\frac{J_1^A}{2(1-J_1^A)}$ (setting $J_1^F=1$). 
The numerical studies for small spin ($s$ = 1/2 and 1) show good agreement with this result \cite{sahoo14}. 

For $s>1/2$ and for a given $J_2^F$, the system goes from ferromagnetic to nonmagnetic spiral phase with increasing $J_1^A$. This 
phase transition can be explained in terms of the broken symmetry approach of Landau, and for the AFAF model, this particular phase transition 
has beed studied \cite{hida13,sahoo14}.

\subsection{Spiral phase for $s>1/2$}

\begin{figure}
%\begin{center} \includegraphics[width=7.5cm,height=6.0cm]{str_fac}
\begin{center} \includegraphics[width=\columnwidth]{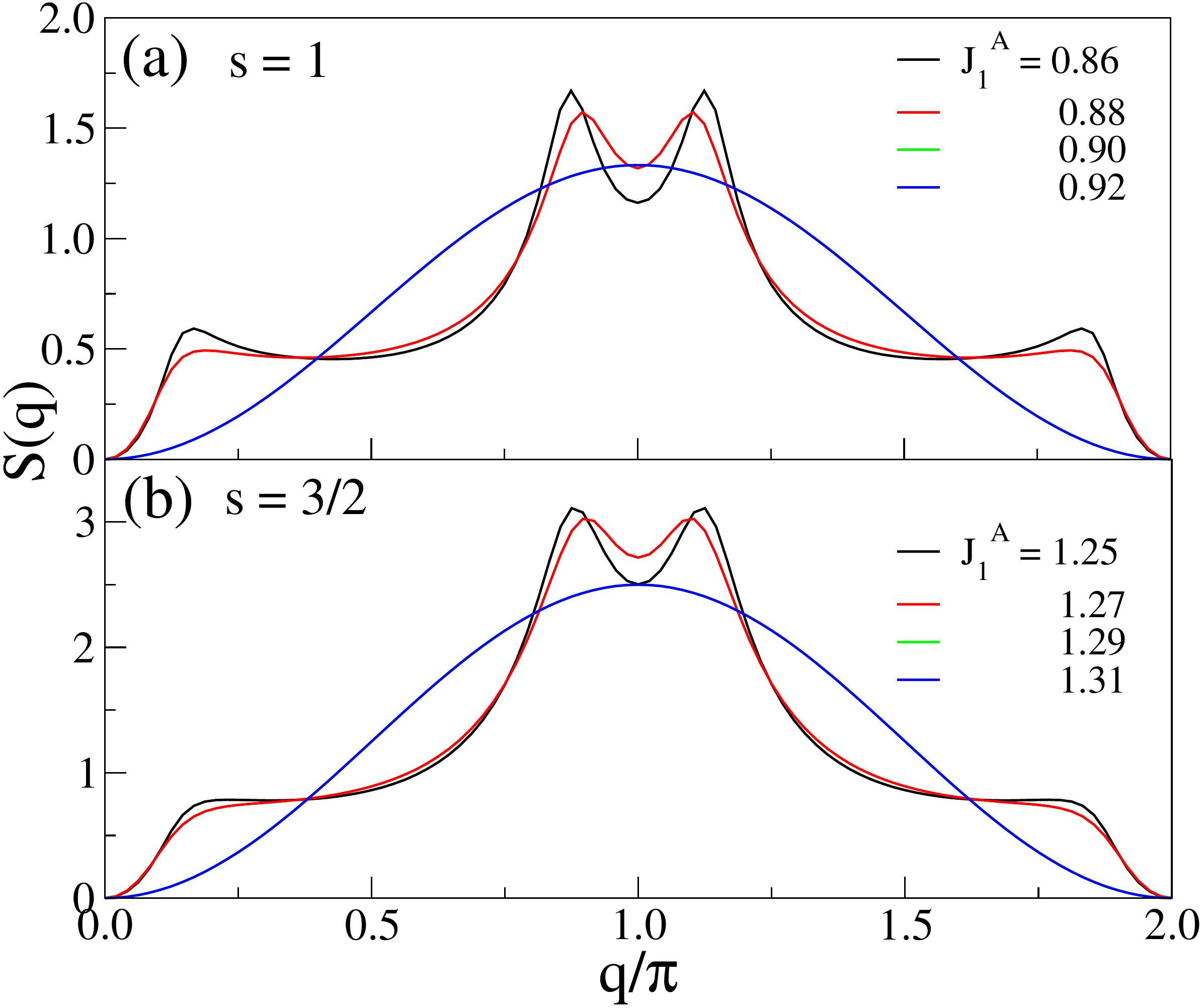}
\caption{Structure factor $S(q)$ is shown for some parameter values across the point $C$ on $J_2^F=1/2$ line. Both for $s=1/2$ and 3/2, the
calculations are done for open chain of length $N=96$.  
}
\label{strfc}
\end{center}
\end{figure}

We already discussed that, our system is in Dimer phase along $J_2^F=1/2$ line as long as $J_1^A>J_C$. To know the nature of the phase 
for $J_1^A<J_C$, we do the structure factor analysis across the point $C$ for both $s=1$ and 3/2. We calculate the structure factor, $S(q)$, 
in the following way: $S(q)=\frac{1}{N} \sum_{l,m} <s_l^zs_m^z> {\rm exp(-iqr_{lm})}$, where $<s_l^zs_m^z>$ is the correlation between 
the $z$-components of spins at sites $l$ and $m$, and $r_{lm}$ is the distance between two sites ($r_{lm}=m-l$). Here $N$ is the total number 
of spins in the system; for calculation of $S(q)$, we consider here an open chain with $(\frac N 2)$-th site as the reference. 
By convention the range of the wave vector is taken as: $0\le q <2\pi$. The $q$ value corresponding to the maximum of $S(q)$ gives us the 
information about the ``spin orientation" in a particular quantum phase (for us $S(q)$ is always positive). We see from FIG. \ref{strfc} 
that, for both $s=1$ and 3/2, $q_{max}$ is close to but less than $\pi$ when $J_1^A<J_C$ ($q_{max}$ is the $q$ value corresponding to 
the maximum of $S(q)$). This implies that we have a spiral phase below the point $C$ in the phase diagram FIG. \ref{phsdiag}. For 
$J_1^A>J_C$, $S(q)$ shows a broad peak at $q=\pi$ -this finding is consistent with our result that in the said parameter
regime we have a Dimer phase which has short range correlation. After checking for some representative points in the phase diagram, we 
find that this spiral phase exists in the whole region between the ``ABCD" line and the ferromagnetic phase. The said ``ABCD" 
line lies inside the NM phase and it separates the spiral phase from the Haldane-Dimer phase. The character of the ``ABCD" line is discussed 
in the next section.  

It may be mentioned here that this spiral phase does not appear for $s=1/2$, for which the point $C$ falls in the FM-NM 
boundary. With increasing value of $s$ ($>1/2$), the location of the point $C$ is expected to go up (here we may remember that the calculated 
upper bound of $J_C$ is $s$ without magnetic field). Our numerical results for $s=1$ and 3/2 support this expectation. 
So we conjecture here that this spiral phase exists for all $s>1/2$. Next we
will see that this particular spiral phase is topologically different than the Haldane-Dimer phase.

\section{Topological phase boundary}
\label{sc5}

For $s>1/2$, a spiral phase appears between the ferromagnetic (FM) and the Haldane-Dimer (HD) phases. The point $C$ lies on 
the phase boundary between the FM and HD phases (FIG. \ref{phsdiag}b). To understand the nature 
of the phase boundary, we study the string-order parameter and the spin-gap across the boundary. 

As discussed earlier, the string-order parameter $O_S$, as defined in Eq. \ref{sopmr}, accurately identifies for our spin model 
whether a phase has topological character or not. We inferred this from the fact that, $O_S$ is nonzero for the half-odd integer dimer state 
(which is topologically equivalent to an SPT phase) and $O_S$ is zero for the integer dimer state (which is equivalent to even spin trivial Haldane
phase).

To characterize the nature of the phase bounday (``ABCD" line as appear in FIG. \ref{phsdiag}b), we first note that the system goes 
gapless at point $C$; this is verified for $s=1$ and 3/2 within our numerical accuracy. The relevant results on the spin gap for $s=1$ and $s=3/2$ can 
be found in FIGs. \ref{spn1plts}a and \ref{spn_intby2}b respectively.

We next study the string-order parameter ($O_S$) across the point $C$ along the $J_2^F=1/2$ line in the phase diagram. Above the point $C$,
where the NN dimer state is the groundstate, $O_S$ = 0 or 1/4 depending of whether $s$ is an integer or half-integer (see earlier discussion). 
Below this point $O_S$ takes some other nonzero value (calculated numerically for $s=1$ and 3/2). We see a sudden change in $O_S$ at the point $C$.
The relevant results for $s=1$ and 3/2 can be seen in FIGs. \ref{spn1plts}b and \ref{spn_intby2}b respectively. This sudden change in $O_S$ 
indicates that, the system goes through a first order topological phase transition at point $C$. But it may not be usual one, since the 
spin-gap at the point $C$ vanishes, which generally indicates a second order quantum phase transition. In fact, the transition point $C$ 
can also be viewed as the symmetry breaking phase transition point as it seperates a dimer and a spiral phase. This type of concurrence
of topological and symmetry breaking phase transitions is not new as mentioned in the introduction of this paper. It is also worth noting that, 
both for integer and half integer spin cases, the point $C$ appears to be a topological phase transition point although only for half integer 
case the Dimer phase is an SPT phase. A more extensive study is needed to have better understanding on this issue.

A topological phase transition point can not be an isolated point inside the phase diagram, as otherwise a particular phase can be shown to 
have two different topological characters. We therefore expect, for our
frustrated AFAF spin model, the point $C$ to lie on a topological phase transition line which spreads 
across the phase diagram. To confirm that the line (``ABCD" in FIG. \ref{phsdiag}b) is actually a topological phase transition line, 
we study $O_S$ across some representative points along the line. For example, when $J_2^F=0.4$, we find that the change in 
$O_S$ across the line becomes sharper with the increasing system size (see inset of FIG. \ref{spn1plts}b). This suggests that the ``ABCD" 
is a topological phase transition line which separates the Haldane-Dimer phase from the spiral phase.

\section{Concluding remarks}
\label{sc6}
In this paper we study the topological aspect of AFAF model for general spins. 
In two opposite limits of $J_1^F/J_1^A$, this model gives two phases - the spin-$2s$ 
Haldane and Dimer, where latter one results from singlet formation between two neighboring spin-$s$ objects.
To have broader understanding of the physics of the model, we study it with 
frustration which is induced by the NNN ferromagnetic interactions ($J_2^F>0$). For the frustrated AFAF model,
the NN dimer state is shown to be the exact groundstate provided $J_1^A \ge J_C$ and $J_2^F=1/2$. 
We also show that the frustrated spin model for $s=1/2$ and $s>1/2$ behave differently: while in the first case the phase 
diagram consists of a ferromagnetic and a nonmagnetic (Haldane-Dimer) phase, in the later case, the model additionally shows a nonmagnetic gapped 
spiral phase which comes between the Haldane-Dimer and ferromagnetic phases. 

The study of a suitably defined string-order parameter and spin-gap at 
the phase boundary indicate that the boundary seperating the Haldane-Dimer and spiral phases is a topological phase transition line 
both for integer and half-integer spins although only half-integer Dimer phase has nontrivial topology. The spiral phase for both types of 
spins appears to have nontrivial topological order. Interestingly,
our studies indicate that the phase boundary line can also be viewd as the second order symmetry breaking phase transition line. 
A more detailed study on the spin model is needed in future to gain better understanding on this issue.

The present work sheds some light
upon the intricacies related to the Haldane physics and opens a new avenue to investigate the many
body topological phases. The AFAF model has already been realized in many systems, and this work may excite the experimentalists to design new 
compounds and study topological phase transitions.

\section*{Acknowledgments} 
SS and DD thank SNBNCBS for supporting them under EVLP. MK thanks the Department of Science and Technology, India and 
SKS thanks DST-INSPIRE for financial support. 

\setcounter{equation}{0}
\setcounter{figure}{0}
\setcounter{table}{0}
\renewcommand{\theequation}{A\arabic{equation}}
\renewcommand{\thefigure}{A\arabic{figure}}

\section*{APPENDIX: NUMERICAL METHODS}

We have used the density matrix renormalization group (DMRG) method, which is a powerful numerical technique
for studying 1D and quasi-1D systems~\cite{white92,white93}. In this technique the truncation of the irrelevant
degrees of freedom is done systematically. For the calculations in this article, we employ
periodic boundary condition (PBC) and for that we use a recently developed efficient DMRG algorithm for systems
with PBC \cite{dey16}. In this algorithm we start with a superblock that consists of eight sites: two sites
in the left and the right block, and two new sites at both the ends of both the blocks. The left and right
blocks increase by four sites as two new sites are added at both the ends of each block. In this way we avoid
the long bond between the old blocks. Here, we have kept up to $m = 500$ eigenvalues of the density matrix to
keep the largest truncation error below $10^{-11}$.

The spin gap $\Delta_S$ is defined as the difference between the singlet groundstate and the triplet first
excited state:
\begin{equation}
\Delta_S (N) = E_0(S^z=1, N) - E_0(S^z=0, N),
\end{equation}
where $E_0(S^z=1, N)$ is the lowest energy in the total $S^z = 1$ sector i.e., the lowest triplet energy and
$E_0(S^z=0, N)$ is the lowest energy in the total $S^z = 0$ sector i.e., the singlet groundstate energy for
a ring of $N$ spins. Calculation of $\Delta_S$ using DMRG is straight forward as our algorithm uses $U(1)$
symmetry to conserve the total $S^z$.

The string order parameter, $O_S$, used in this work is defined in Eq. \ref{sopmr}.
In our calculations, we consider FM interaction $J_1^F$ between the new sites at the both ends of the left or right block.
The left  block is numbered from $2k+1$ to $2l$ whereas, $2k-1$, $2k$, $2l-1$, $2l$ are the new sites.

%%%%%%%%%%%%%%%%%%%%%%%%%%%%%%%%%%%%%%%%%%%%%%%%%%%%%%%%%%%%%%%%%%
%																 %
%																 %
%		REFERENCES											     %
%																 %
%																 %
%%%%%%%%%%%%%%%%%%%%%%%%%%%%%%%%%%%%%%%%%%%%%%%%%%%%%%%%%%%%%%%%%%

\end{document}